\documentclass[pss]{wiley2sp} 
\usepackage{amsmath}
\usepackage{url}

\begin{document}

\title{DFT investigations of the piezoresistive effect of carbon nanotubes for sensor application}

\titlerunning{DFT investigations of the piezoresistive effect of CNTs}

\author{%
  Christian Wagner\textsuperscript{\Ast,\textsf{\bfseries 1}},
  J{\"o}rg Schuster\textsuperscript{\textsf{\bfseries 2}} and
  Thomas Gessner\textsuperscript{\textsf{\bfseries 1,2}}
}

\authorrunning{Christian Wagner et al.}

\mail{e-mail
  \textsf{christian.wagner@zfm.tu-chemnitz.de}, Phone:
  +49-371-531 38699, Fax: +49-371-531 838699}

\institute{%
  \textsuperscript{1}\,Center for Microtechnologies, Chemnitz University of Technology, 09126 Chemnitz\\
  \textsuperscript{2}\,Fraunhofer Institute for Electronic Nanosystems (ENAS), Technologie-Campus 3, 09126 Chemnitz
}

\hyphenation{nano-tubes}

\received{30 April 2012, revised 7 September 2012, accepted 14 September 2012} 
\published{29 October 2012} 


\abstract{%
%
%
%
\abstcol{%
  We investigate the piezoresistive effect of carbon nanotubes (CNTs) within density functional theory (DFT) aiming at application-relevant CNTs. CNTs are excellent candidates for the usage in nano-electromechanical sensors (NEMS) due to their small band gap at zero strain leading to a finite resistivity at room temperature. The application of strain induces a band gap-opening leading to a tremendous change in the resistivity. DFT with the LDA approximation yields reasonable results for pure 
  }{%
    carbon systems like CNTs and is applied to calculate the electronic structure of experimentally relevant CNTs. For the transport part, a simple ballistic transport model based on the band gap is used. We compare our DFT results for the band gaps of strained CNTs to results of tight binding (TB) models. By introducing a scaling factor of $\sqrt{2}$, an excellent agreement of the the DFT data with TB model published in \cite{Yang_2000} is obtained.}}

%
%

\maketitle   

\section{Introduction}
Carbon Nanotubes (CNTs) are very interesting systems from both, the physical and the technological, point of view. In addition to their outstanding mechanical properties like high stiffness and an enormous Young's modulus of about 1 TPa \cite{Wu_2008}, their electronic structures show a rich variety of interesting effects.

One of them is the huge piezoresistive effect (up to one order of magnitude per percent strain), which is the change of the resistivity due to a deformation, relying on the change of the band gap. This behavior is highly desirable for NEMS sensor applications. The effect is strongly dependent on the chiral angle -- in the same way the electronic structure depends on the chirality. Therefore, atmoistic models are needed to build up device simulations.

The modeling of the piezoresistive effect of CNTs using TB models is pioneered by Yang and Han \cite{Yang_2000} as well as Kleiner and Eggert \cite{Kleiner_2001}. The latter one tries to include curvature effects of CNTs by introducing another TB constant. Because its corrections are made using curvature effects, this model is most likely applied for semimetallic CNTs, which have a tiny band gap. The approximations made in these models are not yet fully confirmed by a more advanced theory e.g. density functional theory (DFT).

TB theory only assumes nearest-neighbor coupling of localized electronic states. Therefore, coupling of opposite carbon atoms in the tube is not straight forward, though it is possible to find a corrected band gap formula used in \cite{Kleiner_2001} being valid for small-gap CNTs (where $[n-m]_3=0$). However, these effects may introduce more additional features in the band structure, which are not covered by a TB model. They may be important for transport calculations, because not only the band gap, but also other details like effective mass and higher-order bands are important for charge transport. Therefore, it is beneficial to use DFT as a benchmark.

Since the detailed geometry of a CNT determines its electronic structure very sensitively, the understanding of mechanics is explored in advance. The obtained results are compared to literature data. In a next step, electronic properties are investigated. Between the different models, we find deviations that need explanation.

\section{Calculation details}

Atomistix ToolKit (ATK) from QuantumWise is used \cite{Quantumwise_2011,Brandbyge_2002} to calculate the electronic structure of the carbon nanotubes within the DFT framework. For the exchange functional, local density approximation (LDA) is chosen, because it is the one with the lowest computational cost providing sufficient accuracy for pure carbon systems like CNTs \cite{Tournus_2005}. The Perdew-Wang version of this functional has been applied \cite{Perdew_1991}. The basis set is a double-zeta-polarized (DZP) one, which is recommended for the simulation of CNTs \cite{Abadir_2009}. The CNTs are considered as geometry optimized, when the remaining forces fall below 0.01\,eV/\AA{}. Geometry optimization is performed at each mechanical stretch step induced by linear scaling of the atomic unit cell (including the atoms) along the tube-axis. 

For the calculations of the nanotubes, periodic boundary conditions are used and the k-grid contains (1x1x20) points (Monkhorst-Pack grid). CNTs are strained and compressed linearly. No buckling is observed due to the periodic boundaries, which are cutting off long-range deformations. Geometry optimization gives insight into a realistic structure of the deformed CNTs providing the Poisson's ratio. 

At each deformation state, the band structure (resolved by 201 k-points) is evaluated, analyzed and compared to other data with partly different approaches \cite{Yang_2000,Kleiner_2001,Sreekala_2008,Valavala_2008}. 

\section{Results} \label{sec:results}

\subsection{Mechanical results}

\begin{figure}[t]%
\includegraphics*[width=0.48\textwidth]{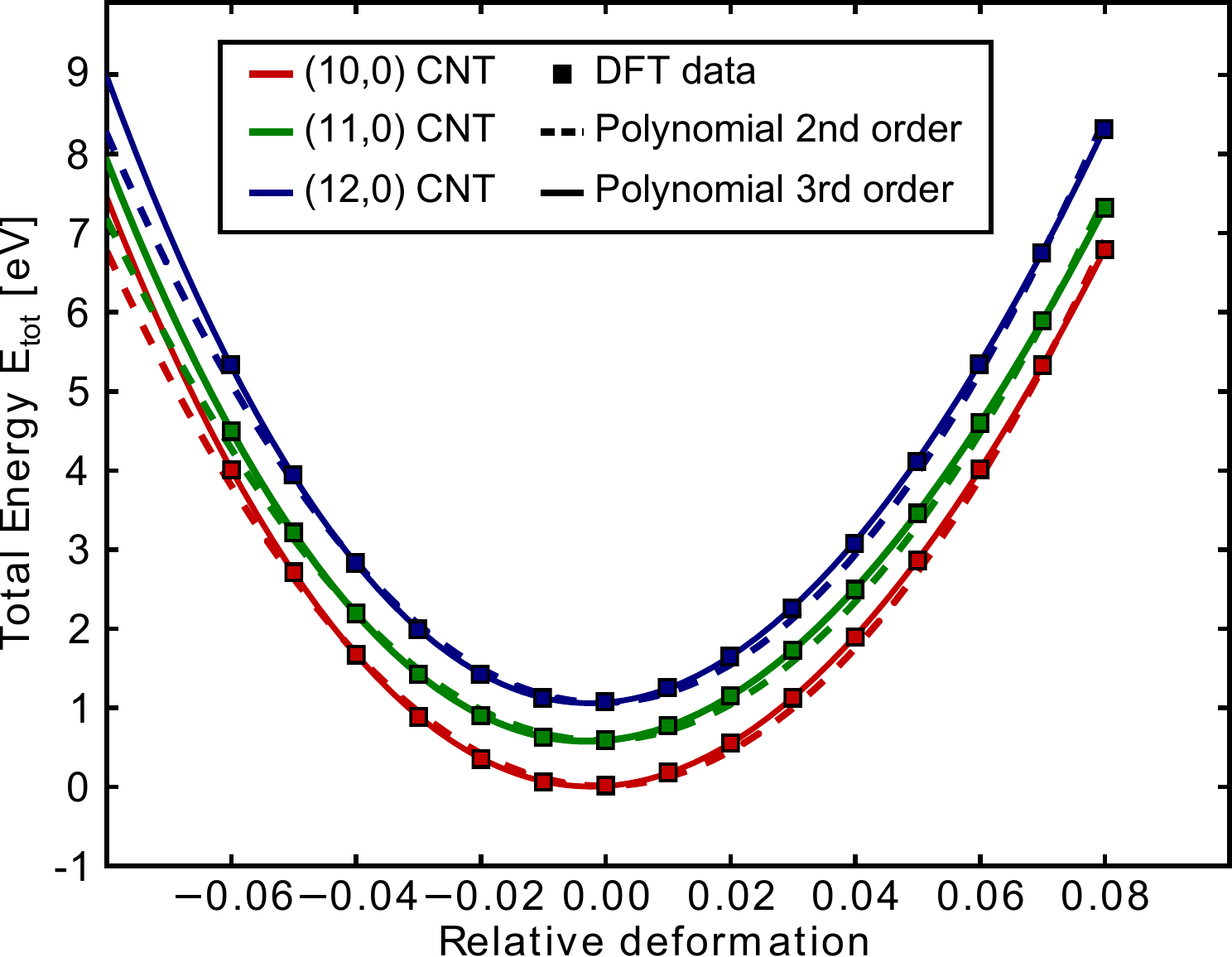}
\caption{The total energy of a (10,0), (11,0), and (12,0) CNT under strain. A third order polynomial fits the data reasonably over the whole deformation range. The data for the different CNTs are shifted by 1\,\textnormal{eV} for clarity.}
\label{fig:total_energy}
\end{figure}

The mechanical data is presented in figure \ref{fig:total_energy}. It shows the total energy of three different zigzag-CNTs (10,0), (11,0) and (12,0) to illustrate the dependency of the total energy on the strain. It can be seen that a third order polynomial is sufficient to describe the data over the whole deformation range allowing a simple extraction of the Young's modulus. Besides the Young's modulus, nonlinear moduli are proposed in literature \cite{Shodja_2011,Colombo_2011} and therefore, $\eta$ from \cite{Shodja_2011} has been determined. The values are shown in table \ref{tab:young}. The Young's moduli are in good agreement with the values of about 1000\,GPa often found in literature \cite{Wu_2008}. In addition, the data fully agree with \cite{Valavala_2008}, where a DFT plain wave approach was used in order to calculate the mechanical and electronic structure of carbon nanotubes. 

More details about the modeling of the mechanical data, the range of the linear regime for those sensor-relevant CNTs can be found in \cite{Wagner_2012}. 

\begin{table}[t]
\centering
  \caption{The Young's modulus as well as the nonlinearity ($\eta$ as introduced in \cite{Shodja_2011}) of different CNTs obtained coefficients of the polynomial fit in figure \ref{fig:total_energy}.}
  \begin{tabular}[htbp]{@{}ccc@{}}
    \hline
    CNT & Young's modulus [GPa] & $\eta$ [GPa]\\
    \hline
    (10,0)  & $976.8 \pm 0.6$ & $980.0 \pm 30.0$\\
    (11,0)  & $977.5 \pm 0.4$ & $978.0 \pm 27.0$\\
    (12,0)  & $981.50 \pm 0.25$ & $982.0 \pm 15.0$\\
    \hline
  \end{tabular}
  \label{tab:young}
\end{table}

\subsection{Electronic structure}
The electronic structure of deformed carbon nanotubes has already been analyzed based on TB models \cite{Yang_2000,Kleiner_2001}. From the application point of view, one can ask if these approaches provide realistic data in combination with a ballistic transport model which is often used \cite{Minot_2003,Cullinan_2010}. Therefore it is examined, if those models hold against DFT calculations. Before doing this, a detailed understanding of the models is required.

\begin{figure*}[t]%
\includegraphics*[width=0.48\textwidth]{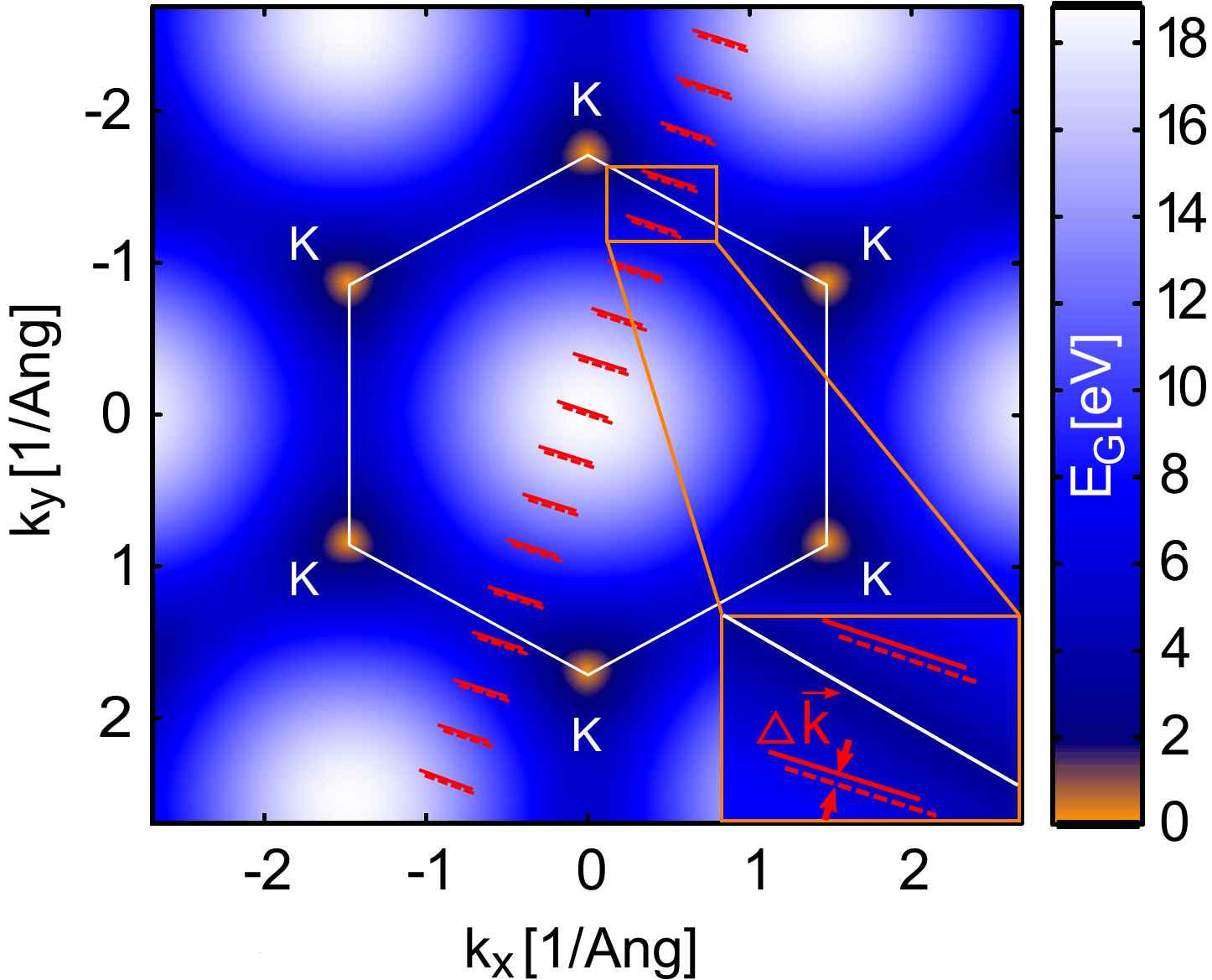}\hfill
\includegraphics*[width=0.48\textwidth]{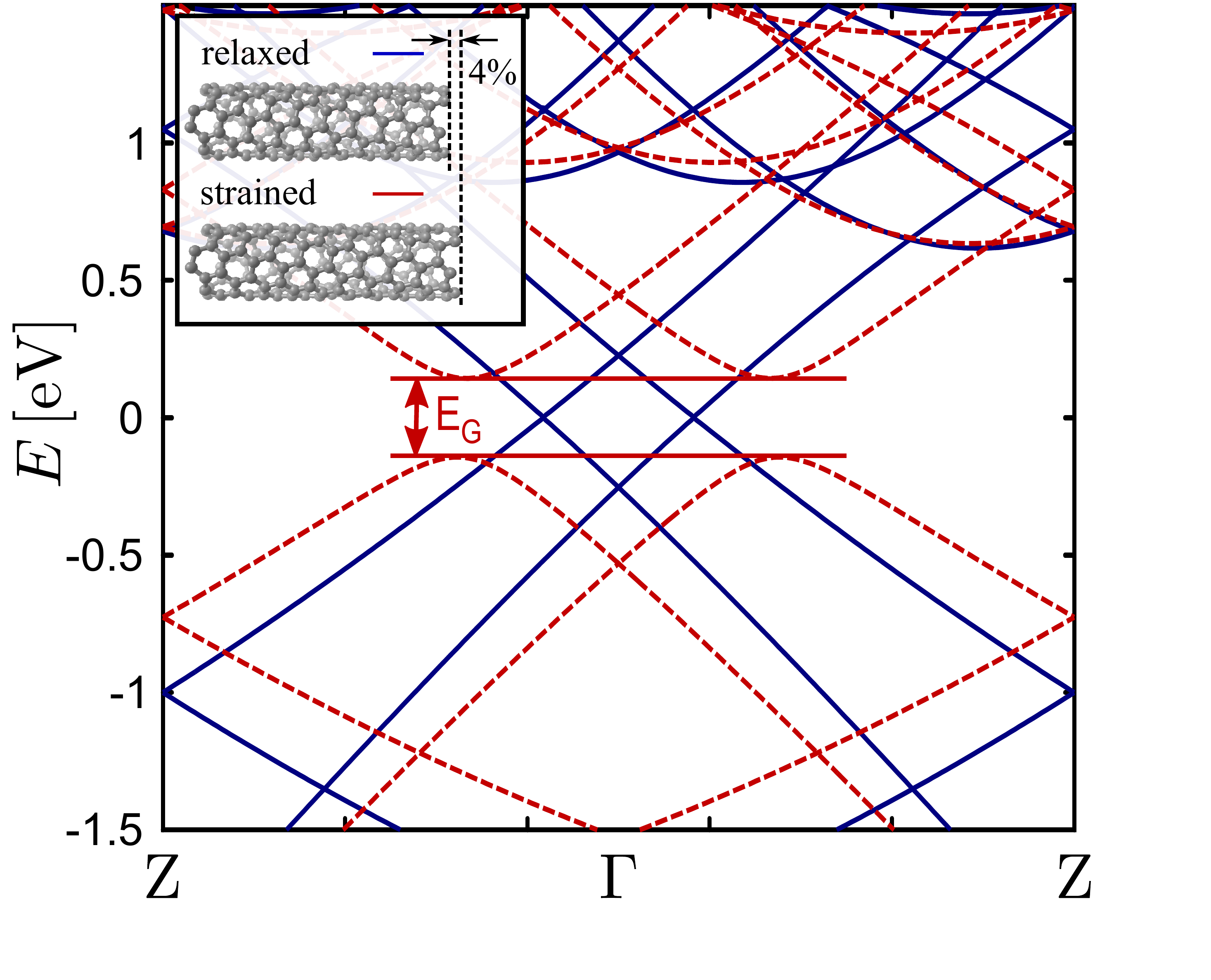}
\caption{The Brillouin-zone (left) and the band structure (right) of the (6,3)-CNT without (straight lines) and with applied strain (dashed lines). The shift $\Delta\vec{k}$ of the k-lines relative to the k-points (left) is responsible for the observed bandgap opening, in the right figure.}
\label{fig:k_space}
\end{figure*}

The theory in \cite{Kleiner_2001} is more formal than \cite{Yang_2000} and also includes curvature effects in an empirical way, which are known to generate tiny band gaps of semimetallic CNTs. The approach of \cite{Yang_2000} is presented in figure \ref{fig:k_space}: The left side of this figure shows the k-space of graphene with the according k-lines representing the periodic boundary conditions in a (6,3)-CNT. The color indicates the difference between the valence- and the conduction band of graphene in the k-space according to the TB formula derived in \cite{Wallace_1947} with the parameters $s=0.13$ and $t_0=2.66\,$eV. $t_0$ is taken from \cite{Yang_2000} and $s$ introduces the asymmetry in the bandstructure of graphene within DFT data (e.g. shown in \cite{Reich_2002}). These k-lines are shifted by the depicted vector $\Delta \vec{k}$ \cite{Yang_2000} due to 4\% strain, which is the consequence of the coordinate transformation imposed by the deformation. This shift influences the band structure strongly, as it is illustrated in the right figure, which represents DFT results. The pristine (6,3)-CNT is shown by straight lines and the strained state by the dashed red lines. The deformation gives rise to a band gap opening, leading to an enormous increase of the resistivity of the CNT. 

\begin{figure*}[t]%
\includegraphics*[width=0.48\textwidth]{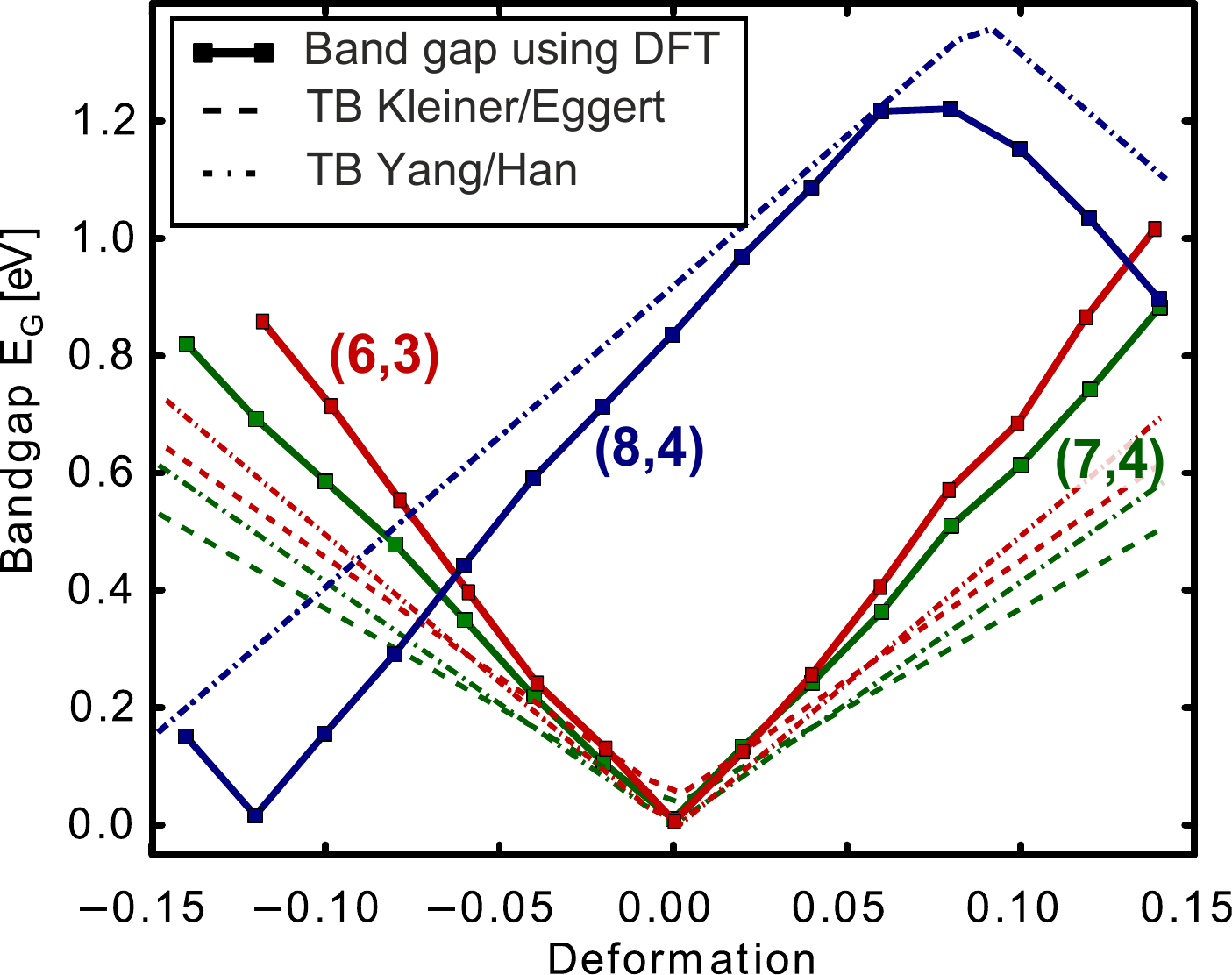}
\hfill
\includegraphics*[width=0.48\textwidth]{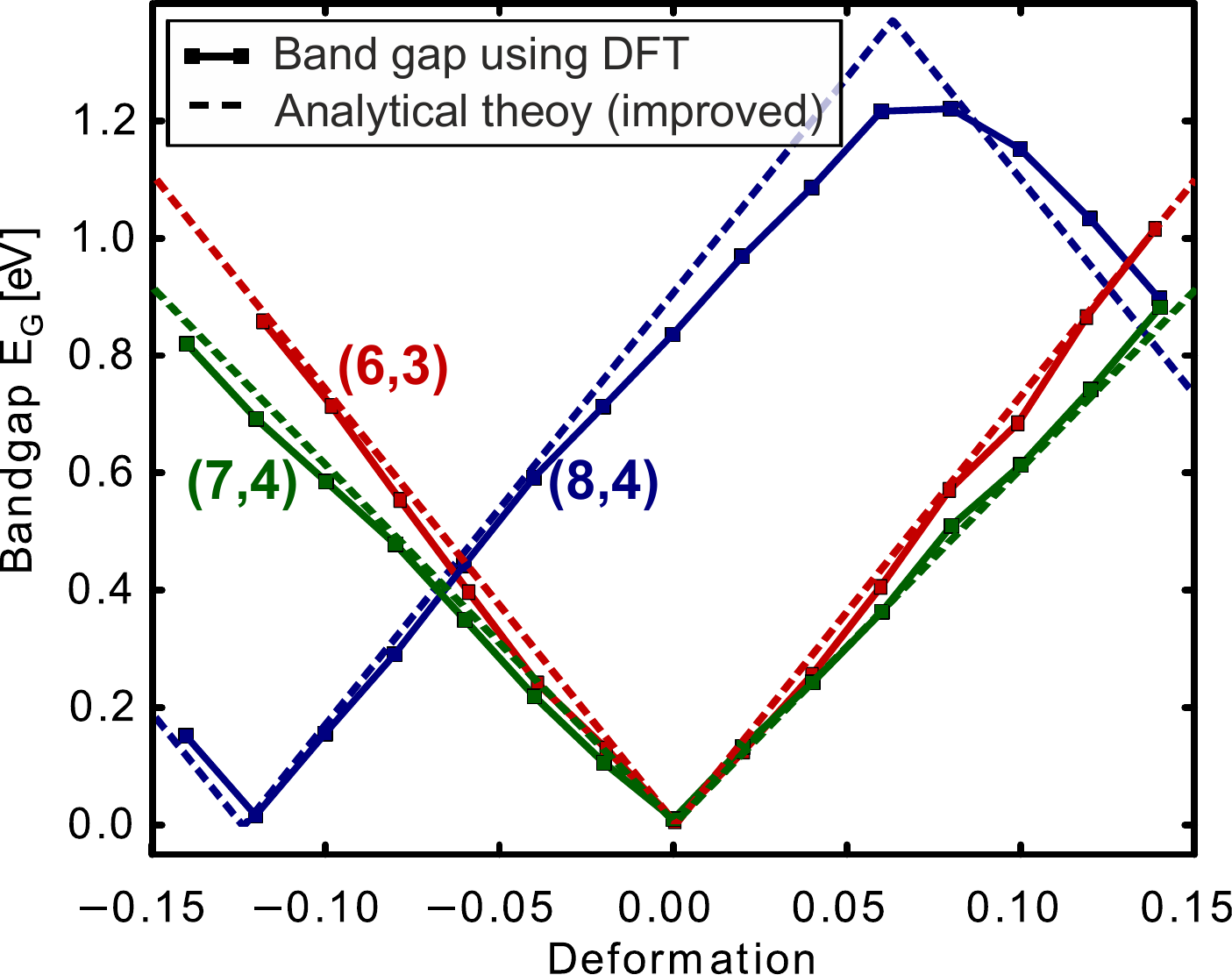}
\caption{Comparison of the DFT band gap with the analytical approaches (left): Dots represent DFT data, straight lines are a guide to the eye, and dashed lines represent the theory by \cite{Kleiner_2001}. Dash-dotted lines are obtained from the theory in \cite{Yang_2000}. Differences between the analytical approach in \cite{Yang_2000} and DFT can be overcome by introducing a factor of $\sqrt{2}$ in the analytical formula (right).}
\label{fig:band gaps}
\end{figure*}

This behavior can now be traced over the whole deformation range by subsequent stretch steps. The resulting band gaps of three different CNTs are shown in figure \ref{fig:band gaps} (left) in comparison to the available theories \cite{Yang_2000,Kleiner_2001}. Dots represent DFT data, straight lines are only to guide the eyes, and dashed lines represent the theory by \cite{Kleiner_2001} as well as dash-dotted lines are obtained from the theory in \cite{Yang_2000}. It can be seen that there is a significant deviation of the DFT data and the analytical formula. A similar difference has been observed in literature for the (10,0) CNT \cite{Valavala_2008}, where our DFT results completely agree with the published ones despite of the different DFT approach. Furthermore, our work agrees well with another work presenting DFT data \cite{Sreekala_2008}. The remaining differences in the compressive part can be attributed to the coarse k-grid-mesh (1x1x4 points), there.

The main impact on the ascend of the band gap-strain relation is the relative shifting of the k-lines to the K-points in the Brillouin-zone. As the k-lines are close, the band gap-opening (and also closing) due to strain is affected by the dispersion relation of graphene in the vicinity of the K-points. There, DFT band structure, TB results, and the common quantification of the linear dispersion ($v_F$, Fermi velocity) agree nicely, as can be seen e.g. in \cite{Reich_2002}. This is also true for our DFT calculations. Thus, the difference of the DFT and the analytical models remains unclear.

For a better understanding of this difference, the zone-folding scheme is applied to calculate the electronic structure in combination with the shift of the k-lines from \cite{Yang_2000}. In order to reproduce our DFT results, this shift has been scaled by a factor of $\sqrt{2}$. This procedure yields a perfect agreement for all CNTs studied so far. Thus, we entered the $\sqrt{2}$ into the analytical model of Yang and Han:
\begin{equation}
 \Delta E_\textnormal{gap} = \textnormal{sgn}(2p+1)\sqrt{2}\cdot 3t_0\,\left[\,(1+\nu)\,\sigma \cos 3\theta \,\right]\ ,
\end{equation}
where $p=[n-m]_3$ with $p=2\mapsto p=-1$, $t_0$ is the hopping-parameter, $\nu$ stands for the Poisson's ratio, $\sigma$ denotes the strain, and $\theta$ represents the chiral angle. This equation holds as long as the same k-line stays the closest one to a K-point. As we did not consider torsional strain, the $\gamma$ in the second term of the original equation has been neglected.

The fact that such a deviation has not been detected so far is most likely due to a lack of systematic comparison of TB results to reference data, which might be either obtained by DFT or experiment. From the experimental side, e.g. in \cite{Leeuw_2008}, data quality is not good enough to precisely determine the ascend of the band gap-strain relation. The difference to DFT data is not that obvious in the literature found \cite{Sreekala_2008,Valavala_2008} and this misfit was either overseen or being attributed to the general known weaknesses of DFT, when band gaps are predicted. However, band gaps are usually underestimated \cite{Mattsson_2005} by DFT which is not the case in the present study. 

All the mentioned facts make us confident that our DFT results are reliable and they are able to correct the analytical formula published by \cite{Yang_2000}. This factor $\sqrt{2}$ should then also come out of the calculation performed in \cite{Yang_2000}.

The change of the band gap due to strain is very interesting for sensor applications and NEMS-devices because the resistivity $R$ of a CNT is exponentially dependent on the band gap \cite{Minot_2003} in the simplest transport approximation

\begin{equation}
R = R_S + \frac{h}{8e^2}\frac{1}{|t|^2} \left(1+ \exp\left( \frac{E_G}{k_BT}\right) \right),
\end{equation}

where $\frac{h}{8e^2}$ is the quantum resistivity in graphene, $E_G$ stands for the band gap, and $T$ represents the temperature. We did not consider contact effects or scattering, so $|t|^2$ is taken as 1 and $R_S$ equals zero.

It should be noted that the zone-folding approach and the TB scheme presented here only holds for CNTs with a diameter larger than $\approx6$\,\AA{}. For CNTs with lower diameters, the curvature (leading to $\sigma^\ast-\pi^\ast-$hybridization, \cite{Blase_1994}) becomes important, which cannot be easily taken into account in a simple, general model for the electronic structure. DFT is expected to provide reliable results also for that kind of CNTs, but it is not expected to find a simple, analytical expression for this behavior.

\section{Conclusion and outlook}
We presented calculations of sensor-relevant CNTs based on DFT. Our mechanical data show a good agreement with current literature. It is found that DFT data and present analytical models differ significantly and could be brought into agreement by a prefactor of $\sqrt{2}$ in the analytical model.

In future work, the analytical calculation is repeated to watch out for this $\sqrt{2}$. Further on, transport simulations based on the DFT data will be accomplished and realistic (defective) and functionalized CNT will also be taken into account.

\begin{acknowledgement}
This work has been done within the Research Unit 1713 which is funded by the German Research Association (DFG). We gratefully acknowledge the ongoing support by the group of Michael Schreiber (TU Chemnitz).
\end{acknowledgement}

%
\bibliographystyle{pss}
%

\providecommand{\WileyBibTextsc}{}
\let\textsc\WileyBibTextsc
\providecommand{\othercit}{}
\providecommand{\jr}[1]{#1}
\providecommand{\etal}{~et~al.}

\end{document}